# 3D Smith chart constant quality factor semi-circles contours for positive and negative resistance circuits

Victor Asavei[1], Member IEEE, Andrei A. Muller[2], Senior Member IEEE, Esther Sanabria-Codesal[3], Alin Moldoveanu[1], Member IEEE, Adrian. M. Ionescu[2], Fellow IEEE

[1]Department of Computer Science and Engineering, Faculty of Automatic Control and Computers, University Politehnica of Bucharest, 060042 Bucharest, Romania
[2]Nanoelectronic Devices Laboratory (NanoLab), École Polytechnique Fédérale de Lausanne (EPFL), 1015, Lausanne, Switzerland
[3]Departamento de Matemática Aplicada, Universitat Politècnica de València, Valencia, Spain

Corresponding author: A.A. Muller (e-mail: andrei.muller@epfl.ch).

The work was supported by the Phase-Change Switch Fet-Open H2020 grant 737109 and by DGC grant 2018-094889-B-100.

**ABSTRACT** The article proves first that the constant quality factor ($Q$) contours for passive circuits, while represented on a 2D Smith chart, form circle arcs on a coaxal circle family. Furthermore, these circle arcs represent semi-circles families in the north hemisphere while represented on a 3D Smith chart. On the contrary we show that the constant $Q$ contours for active circuits with negative resistance form complementary circle arcs on the same family of coaxal circles in the exterior of the 2D Smith chart. Also, we find out that these constant $Q$ contours represent complementary semi-circles in the south hemisphere while represented on the 3D Smith chart for negative resistance circuits. The constant $Q$ - computer aided design (CAD) implementation of the $Q$ semi-circles on the 3D Smith chart is then successfully used to evaluate the quality factor variations of newly fabricated Vanadium dioxide inductors first, directly from their reflection coefficient, as the temperature is increased from room temperature to 50 degrees Celsius (°C). Thus, a direct multi-parameter frequency dependent analysis is proposed including $Q$, inductance and reflection coefficient for inductors. Then, quality factor direct analysis is used for two tunnel diode small signal equivalent circuits analysis, allowing for the first time the $Q$ and input impedance direct analysis on Smith chart representation of a circuit, including negative resistance.

## I. INTRODUCTION

The Smith chart [1-2], proposed by Philip Hagar Smith in 1939, survived the passing of years, becoming an icon of microwave engineering [3], being nowadays still used in the design and measurement stage of various radio frequency or microwave range devices [4-7], for plotting a variety of frequency dependent parameters.

Constant quality factor ($Q$) representations on the Smith chart are a visual way to determine the quality factor of various passive microwave circuits, being mostly known in the microwave frequency range community [8-17]. These constant Q shapes are often denoted as contours or curves, [8-12] while extremely seldom as circles [11].

In this work we first prove that the constant $Q$ curves represent circle arcs mapped on coaxal circle families [18] while providing to the best of our knowledge for the first time their equations: i.e. centre-radius-$Q$ dependency.

Then, we prove that these circle arcs represent simple semi-circles on the North hemisphere for all passive circuits while analysed on the 3D Smith chart computer aided design (CAD) tool [19-21]. After displaying the constant $Q$-arcs for negative resistance circuits, we determine that these are semi-circles in the south hemisphere.

In order to prove the utility of our CAD implementation, we show that, while grounding the second port of newly fabricated Vanadium dioxide two port inductors [22], one may get the $Q$-frequency dependency directly from the $S_{11}$-reflection coefficient parameter, thus avoiding classical 2D $Q$-frequency plots previously employed by us [22-23] or other authors [24-25], in these types of evaluations. The proposed visualization on the 3D Smith chart proves its effectiveness for passive circuits, especially when quality factors do not exceed big values [22-25], being particularly useful for Vanadium dioxide temperature variations studies on *Q*-where *Q* degrades as temperature increases, but potentially applicable in all inductors frequency dependent factor evaluations. Here we test the temperature dependence of *Q* for fabricated inductors with $VO_2$ by sweeping it from 25°C to 50°C-directly from the vector network analyser with the newly developed technique. Further we show the utility of the new CAD implementation for negative resistance circuits and we analyse the quality factor of various tunnel diodes, when negative resistance occurs, and the Smith chart cannot be used anymore.

## II. Constant Q semi-circles representation and applications

### A. Constant Q circle arcs on the 2D Smith chart-equations,

The quality factor of an impedance *Z* or admittance *Y* can be defined as (1) [12-17], (or denoted nodal quality factor in [15-17] where *X* represents its reactance, *B* its susceptance, *R* resistance and *G* conductance as defined in (1) where *Z* and *Y* can be related through (2). Other authors skip the absolute value sign in (1) [8-11], however these changes nothing in respect to their geometry, only to the sign labelling convention.

Using the sign conventions [12-17] (as for example at p.102 in [13]) we do not allow negative *Q* values for passive circuits with positive resistances.

$$Q = \frac{|X|}{R} = \frac{|B|}{G}; \quad (1)$$

$$Z = \frac{1}{Y} = R + jX = \frac{1}{G+jB} \quad (2)$$

On the other hand, the reflection coefficient of one port network (where $R_1$ is the port resistance, usually 50 Ω) can be defined as:

$$S_{11}(z) = \frac{Z-R_1}{Z+R_1} = \frac{R+jX-R_1}{R+jX+R_1} = \frac{r+jx-1}{r+jx+1} = \frac{z-1}{z+1} = \rho_r + j\rho_i \quad (3)$$

where *r* and *x* denote the normalized resistance, respectively reactance:

$$r = \frac{R}{R_1}; x = \frac{X}{R_1}, z = r + jx \quad (4)$$

and $\rho_r$ and $\rho_i$ denote the real and the imaginary part of the reflection coefficient.

Based on (1) and (4) one can easily obtain *Q* as (5).

$$Q = \frac{|x|}{r} = n \quad (5)$$

If n≠ 0 is constant, we observe that the expression (5) represents radial lines in the normalized impedance (*z* plane). Fig. 1 (a) shows that for |n|=3 we obtain two lines as *r* and *x* are swept from -∞ to+∞. Fig. 1 (b) shows the family of the radial lines obtained for various values of n.

Imposing now *Q*=*n* constant and positive in (3) one gets the contours obtained in [8-17] irrespective of the presence of the absolute value in (1). Using, however inversive geometry theory [18], it can be proven that imposing (5) in (3) the set of radial lines *Q*=*n* constant generates a family of coaxal circles as *r* and *x* are swept from -∞ to +∞.

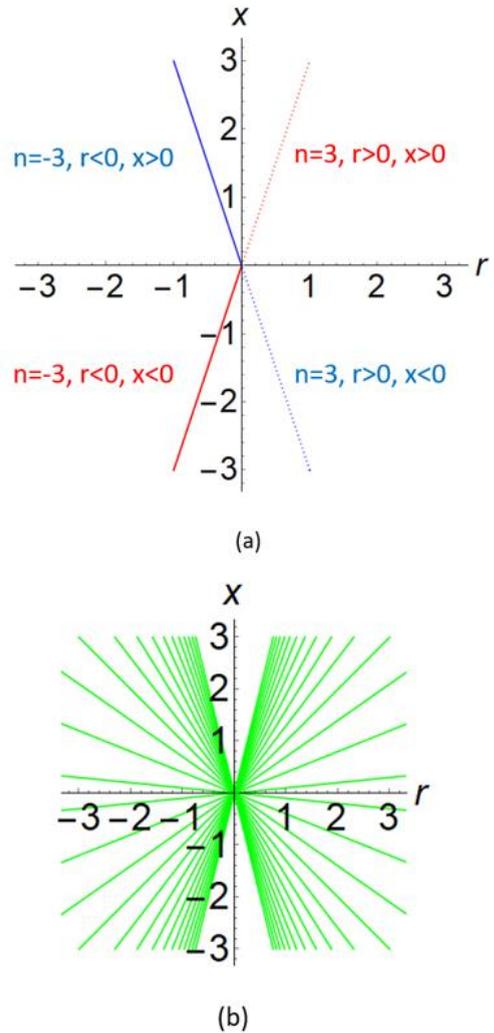

**FIGURE 1.** a) Constant Q=n lines in the z plane a) Q=n=3 blue/red radial lines. As *r* and *x* are swept from -∞ to ∞ |n|=3 for lines passing through 0 and infinity b) -4<n<4.

Proof:
We know from [18-19] that any transformation of form (6) represents an inversive transformation mapping always generalized circles (circles or infinite lines) into generalized circles.

$$A(z) = \frac{az+b}{cz+d} \text{ (a)}; \quad B(z) = \frac{a\bar{z}+b}{c\bar{z}+d} \text{ (b)} \quad (6)$$

Since (3) is a particular case of (6) (Mobius transformation) then, when imposing (5) in (3), the radial lines become generalized circles that passes through the points $S_{11}(z=0) = (-1,0)$, $S_{11}(z=\infty) = (-1,0)$.

By matching the real part of (3) to zero, we assert that the image of the points $z = \frac{1}{\sqrt{1+n^2}} \pm \frac{n}{\sqrt{1+n^2}}j$ determine the cutting points with the $\rho_i$ axes, i.e.

$$S_{11}(z = \frac{1}{\sqrt{1+n^2}} \pm \frac{n}{\sqrt{1+n^2}}j) = (0, \frac{\pm n^2}{\sqrt{n^2+n\sqrt{1+n^2}}}) \quad (7)$$

Then, we obtain a coaxal family of circles that pass through (-1,0) and (1,0), and their centres lie on the $\rho_i$ axes with radical axis $\rho_r$. ([26]).

By using the bipolar coordinates ([27]), we obtain that the centre (*C*) and radius (*rad*) of this family of circles are:

$$C = \left(0, -\frac{sgn(x)}{n}\right), \quad rad = \sqrt{1 + \frac{1}{n^2}} \quad (8)$$

For *n>0* we obtain simply circles arcs inside of Smith Chart as observed too, only in [11].
For *n=0*, we have a circle of infinity radius, i.e. the $\rho_r$ axis. We consider that the distance between two constant *Q*-circles is given by the distance between their intersections with $\rho_i$ axis (7), then the distance between two consecutives circles is given in Table I. Fig. 2 showing the *Q=n>0* circles.

TABLE I
DISTANCE BETWEEN CONSECUTIVE CONSTANT Q CIRCLES (N, N+1) AND REACTANCE AXES CROSSINGS

| N, N+1 VALUES | DISTANCE | INTERSECTIONS WITH $\rho_i$ $\frac{\pm n^2}{\sqrt{n^2+n\sqrt{1+n^2}}}$ |
|---|---|---|
| 0, 1 | 0.414 | 0, ±0.414 |
| 1, 2 | 0.204 | ±0.414, ±0.618 |
| 2, 3 | 0.103 | ±0.618, ±0.721 |
| 3, 4 | 0.06 | ±0.721, ±0.781 |
| 4, 5 | 0.039 | ±0.781, ±0.82 |
| 5, 6 | 0.027 | ±0.82, ±0.847 |
| 6, 7 | 0.020 | ±0.847, ±0.867 |
| 7, 8 | 0.015 | ±0.867, ±0.883 |
| 8, 9 | 0.012 | ±0.883, ±0.883 |
| 9, 10 | 0,009 | ±0.883, 0.905 |

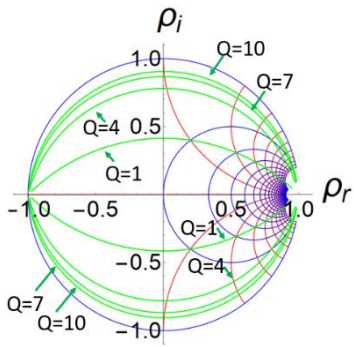

**FIGURE 2.** Various *Q=n* circles arcs inside Smith Chart for n>0.

Computing the distance between two consecutive circles, for negative resistance circuits (with $r \leq 0$ implying $n \leq 0$) we obtain the results in Table II. Displaying now the contours for $n \leq 0$, we get the circle arcs in Fig.3. At the limit, for n=0, we have a circle of infinite radius, i.e. the $\rho_r$ axis. Finally, Fig. 4 displays all the coaxal circles together for -∞≤n≤∞

TABLE II
DISTANCE BETWEEN CONSECUTIVE CONSTANT Q CIRCLES (N, N+1) AND REACTANCE AXES CROSSINGS

| N, N+1 VALUES | DISTANCE | INTERSECTIONS WITH $\rho_i$ $\frac{\pm n^2}{\sqrt{n^2+n\sqrt{1+n^2}}}$ |
|---|---|---|
| -1, 0 | 2.41 | 0, ±2.41 |
| -2, -1 | 0.796 | ±2.41, ±1.61 |
| -3, -2 | 0.23 | ±1.61, ±1.38 |
| -4, -3 | 0.106 | ±1.38, ±1.28 |
| -5, -4 | 0.061 | ±1.28, ±1.22 |
| -6, -5 | 0.0393 | ±1.22, ±1.18 |
| -7, -6 | 0.0274 | ±1.18, ±1.15 |
| -8, -7 | 0.0202 | ±1.15, ±1.13 |
| -9, -8 | 0.0155 | ±1.13, ±1.12 |

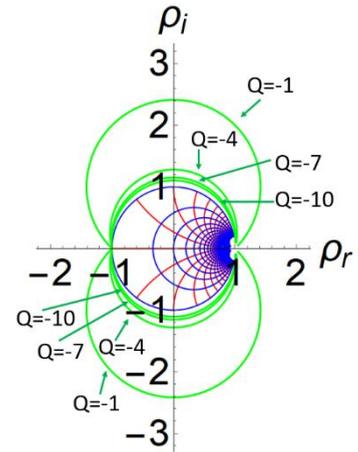

**FIGURE 3.** Various *Q=n* circles arcs in the exterior of the Smith Chart for n<0.

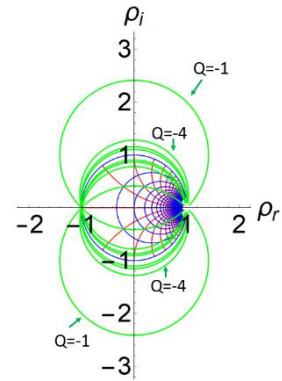

**FIGURE 4.** Various *Q=n* circles for -10<n<10.

*B. 3D Smith chart representation of constant Q semi-circles*

By using the stereographic projection, we can get images of the coaxal circles on the 3D Smith chart [19-22] (9)

$$S_{113D}(S_{11} = \rho_r + j\rho_i) = \left(\frac{2\rho_r}{1+|\rho|^2}, \frac{2\rho_i}{1+|\rho|^2}, \frac{1-|\rho|^2}{1+|\rho|^2}\right) \quad (9)$$

The coaxal circles become semicircles on the 3D Smith chart that pass through the points $S_{113D}(-1,0) = (-1,0,0)$,

$S_{113D}(1,0) = (1,0,0)$ and $S_{113D}\left(0, \frac{\pm n^2}{\sqrt{n^2+n\sqrt{1+n^2}}}\right) = \left(0, \frac{\pm n}{\sqrt{1+n^2}}, \frac{1}{\sqrt{1+n^2}}\right)$.

For $n\neq 0$, these semicircles are given by the intersection of the sphere with the fold plane:

$$-2\frac{\text{Sign}[x]}{n}x_2 + 2x_3 = 0, \qquad (10)$$

where $(x_1, x_2, x_3)$ denote the coordinates on 3D Smith chart, as seen from the centre of it.

If n=0, the corresponding semicircle passes through the points $(\pm 1, 0, 0)$ and $(0, 0, 1)$, i.e. it is Greenwich meridian. Table III summarizes the results generated by (10).

TABLE III
CONSTANT Q CIRCLES SURFACES ON THE 3D SMITH CHART FOR DIFFERENT VALUES OF N

| N | SURFACE | COMMENT |
|---|---------|---------|
| 0 | $x_2 = 0$ (PLANE) | GREENWICH MERIDIAN |
| $-5$ | $2x_3 + \frac{2}{5}x_2 \text{SIGN}(x) = 0$ | POSITIVE FOLD PLANE |
| $+5$ | $2x_3 - \frac{2}{5}x_2 \text{SIGN}(x) = 0$ | NEGATIVE FOLD PLANE |
| $\pm\infty$ | $x_3 = 0$ (PLANE) | EQUATOR OF THE 3D SMITH CHART |

Fig. 5 shows the constant $Q$ circles on the 3D Smith chart together with the cutting planes given by (10).

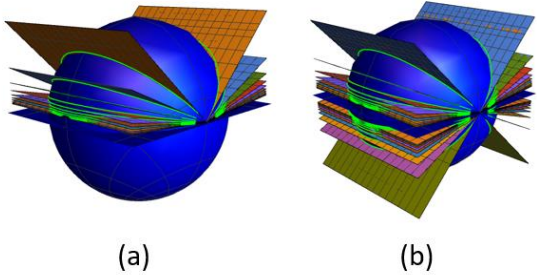

FIGURE 5. *Q=n* semi-circles and their planes on the 3D Smith chart: (a) for r>0 (n>0), (b) for -∞<r<∞, (-∞<n<∞).

A more detailed view is given in Fig. 6 - Fig. 8 for a variety of values of *Q*.

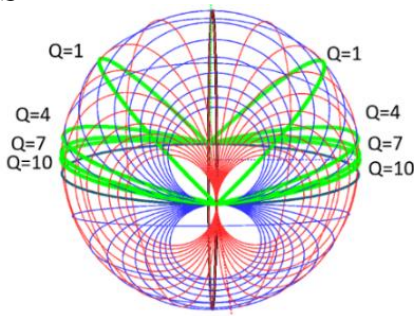

FIGURE 6. Various *Q=n* circles on the 3D Smith chart for 0<n≤10.

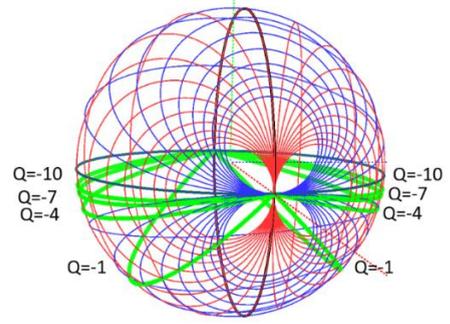

FIGURE 7. Various Q=n circles on the 3D Smith chart for -10<n≤0.

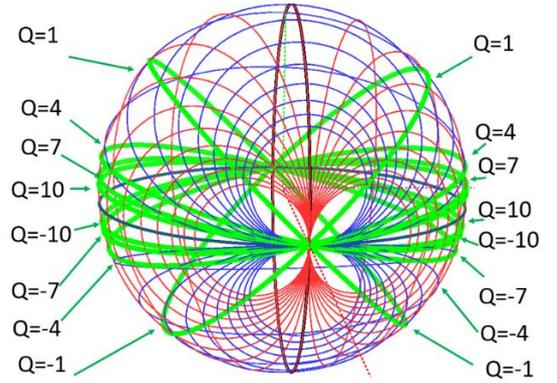

FIGURE 8. Various Q=n circles on the 3D Smith chart for -10≤n≤10.

### C. Applications in reconfigurable inductors with Vanadium dioxide switches in on/off state

The quality factor of an inductor is most commonly defined as (11) - by grounding its second port [23-25, 28-30]. It is usually evaluated until *x* changes sign and becomes negative. Respectively, we may notice that the imaginary part of the reflection coefficient of a one port network can be written as (12) and this becomes zero when *x* becomes 0. Based on this observation, the zeros of the imaginary part of $S_{11}$ are given by the zeros of *x*.

Secondly, (11) is identical with (5) under *r>0* and *x>0* conditions, which are always fulfilled for a passive inductor before its self-resonant frequency (when it becomes capacitive).

$$Q = \frac{Im\left(\frac{1}{Y_{11}}\right)}{Re\left(\frac{1}{Y_{11}}\right)} = \frac{x}{r} \qquad (11)$$

$$\rho_i = \frac{2x}{(1+r^2)+x^2} \qquad (12)$$

Plotting the 1 port reflection coefficient on the Smith chart delivers the values of the quality factor directly, unlike the two port reflection coefficients representation on the Smith chart which is unrelated to it. Displaying the one-port reflection coefficient (grounding the second port), we can directly detect the quality factors and while using the 3D Smith chart implementation, visualize the extracted inductance and frequency dependency in a concomitant view.

Considering our previous work [22], the inductance can be represented in the 3D space via a homothety over the $S_{113D}$ parameter using (13) where $L_N(\omega)$ represents the normalized extracted inductance.

$$L_{3d}(\omega) = (L_N(\omega) + 1) * \rho_{3D}(j\omega) \qquad (13)$$

In [19], the proposed display could not allow for the simultaneous displaying of $L_{3d}(\omega)$, $Q$ and $S_{113d}(\omega)$.

Fig. 9 shows the extracted inductance and $Q$ of the inductors reported in [22] and using VO$_2$ as switching element using a classical approach. The results are measured at 20 °C (off) and at 100 °C (on).

Let us consider now the representation proposed in the previous section. Fig. 10 (a) displays the $S_{113D}$ of the inductors in on/off states, with the second port grounded, for the same frequency range as in Fig.10. One may directly read the value of the quality factor in each point of them using the CAD implementation proposed, while plotting the convenient constant $Q$ semi circles. From the chosen rendering it renders clear that these values are between 7-10 for a wide frequency band, while in the case of the off state inductors, these values are decreasing towards 0. For the on state it can be seen clearly that these values do not decrease below 4 for the frequency range displayed. Fig 10 (b) adds the frequency representation proposed in [22] over the previous render, offering their dynamical view as frequency is swept. Fig 10 (c) shows the extracted inductance displayed over them, using a centre projection for each point. The results picture the increased value of the inductance in the off state as in comparison to the one in the on state, while both showing frequency linearity.

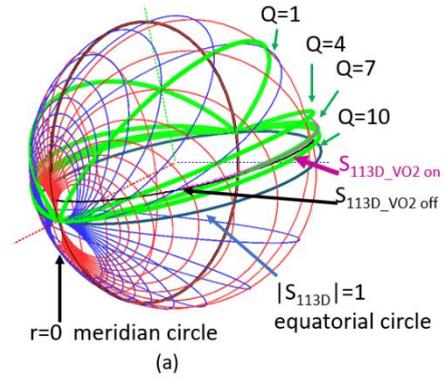

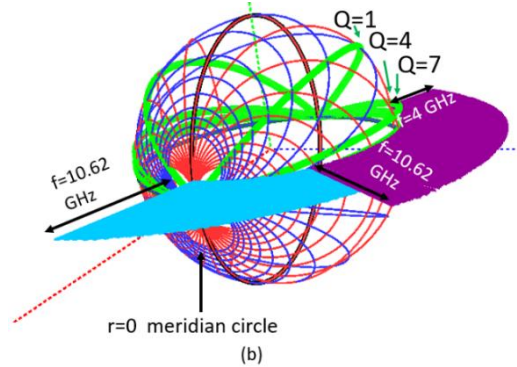

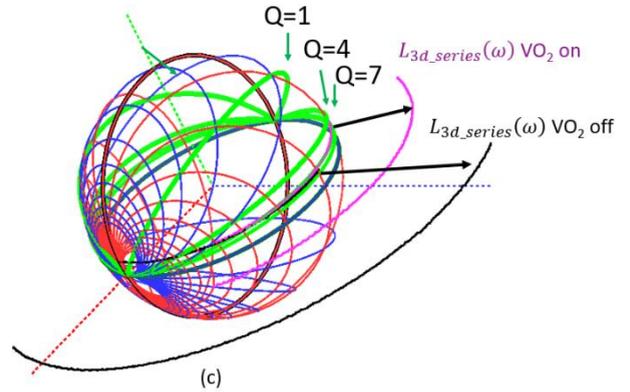

**FIGURE 10.** On (pink) / off (black) frequency dependent extracted parameters of the fabricated inductors between 4 GHz and 10. 62 GHz on the 3D Smith chart(a) $S_{113D}$ (b) $S_{113D}$ and frequency visualization (c) $S_{113D}$ and $L_{3D}$.

### D. Applications in reconfigurable inductors with Vanadium dioxide switches in temperature sweeping

The proposed CAD approach is particularly useful in the case of VO$_2$ inductors, since these need to be tested at various temperatures, thus a fast detection of a failure, directly from the measured S parameters, would allow for skipping of the testing for the same inductor at a different temperature. Usually on a wafer, one which has in the tenths-hundreds number denomination of inductors, the proposed procedure represents a fast tool for detecting failures in the desired expected $Q$. Fig. 11 summarizes the proposed procedure.

The S parameters can be exported as Touchstone two-port files and imported directly into the 3D Smith chart application. The application is developed using the Java programming language and for the 3D rendering and interaction with the Riemann sphere the Open Graphics Library (OpenGL) Application Programming Interface (API) is employed. The user can interact with the 3D space in which the Riemann sphere is rendered to manipulate the view of the 3D space and to adjust the parameters of the displayed circuits, as necessary.

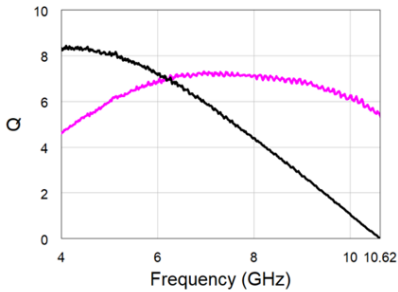

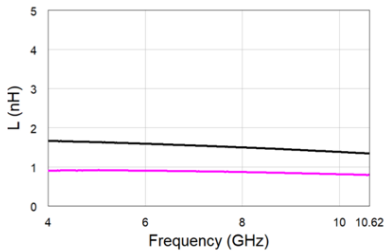

**FIGURE 9.** On, (pink at 100°C) / off (black at 25°C) extracted frequency dependent parameters of the fabricated inductors measured between 4 GHz and 10. 62 GHz (a) Quality factor (b) Inductance.

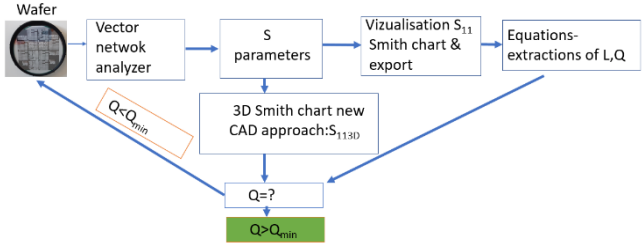

**FIGURE 11.** Proposed CAD Q evaluation, for a wafer full of inductors requiring Q>$Q_{min}$ on a specific frequency band.

In order to test the temperature sensitivity of the Vanadium Dioxide reconfigurable inductors, let us examine a new inductor, based on the design methodology in [19], but with longer switch length (minimizing losses in off state and increasing them in the on state). The measurement setup is shown in Fig. 12 (a)-and it includes a thermo chuck, whose temperature is increased up to 50 °C. Fig. 12 (b) shows the layout of the inductor, the same as in [22], in this case however with a 2 μm switch length instead of 600 nm as in [22].

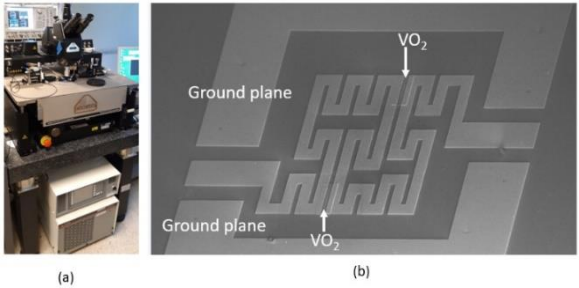

**FIGURE 12.** (a) Setup for heating including Vector network analyser and a thermo heater (b) Layout of the inductors [18].

Let us now verify the $Q$ frequency dependency while analysing the inductors in the 4-8 GHz and then check the minimum value in this band.

The extracted $Q$ and inductance are displayed in Fig. 13 (a) and (b) on a 2D display. The values of the $Q$ decrease slightly up to 6, while the values of the extracted inductance stay stable with temperature increase-Fig 13(b).

Using the new proposed CAD methodology, we can see in Fig. 14 (a), the $S_{113D}(\omega)$. It can be clearly seen how the $S_{113D}(\omega)$ does not decrease below 6 for none of the analysed temperatures.

Exploiting the frequency dependency display, the dynamics in Fig. 14 (b) can be observed. Fig 14 (c) shows the extracted inductance in 3D - displaying its extremely stable values as temperature increases up to 50 °C.

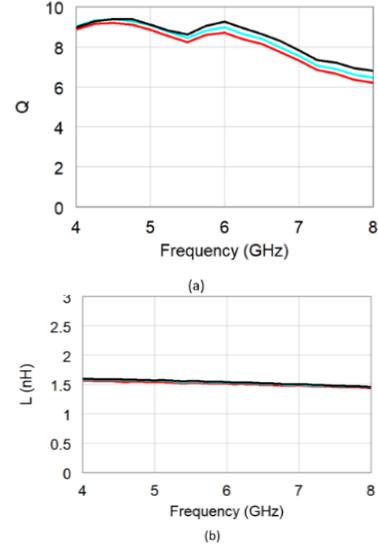

**FIGURE 13.** Temperature dependence of the inductors parameters measured between 4 GHz and 8 GHz while sweeping the temperature from 25 °C (black) to 40 °C (cyan) and 50 °C (red) (a) quality factor (b) extracted inductance.

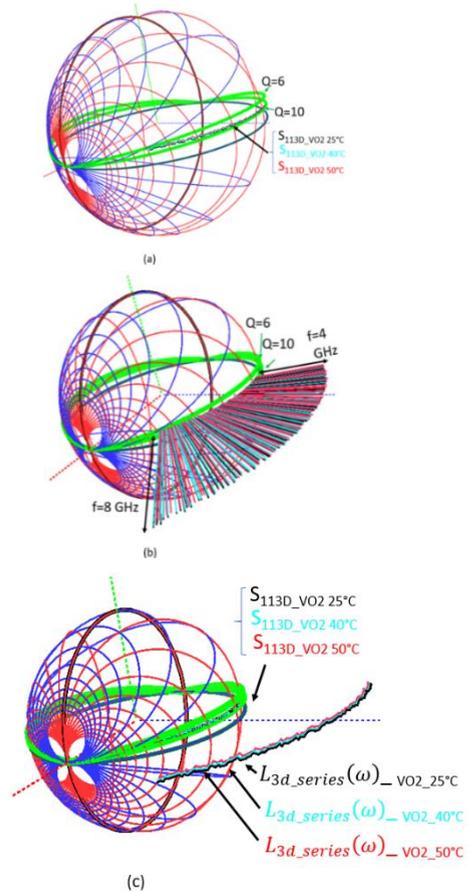

**FIGURE 14.** Temperature dependence of the inductors parameters measured between 4 GHz and 8 GHz while sweeping the temperature from 25 °C (black) to 40 °C (cyan) and 50 °C (red) (a) $S_{113D}$ between the constant Q circles (b) $S_{113D}$ between the constant Q circles including frequency dependency-showing their dynamics $L_{3D}$ (c) $S_{113D}$ and $L_{3D}$ displayed simultaneously.

## III. Negative resistance circuits

### A. Frequency dependent example

Let us consider the circuit given in Fig. 15, which is the equivalent circuit of a resonant tunnelling diode [30]. These diodes can be used as local oscillators in microwave and millimetre wave frequencies. Their small signal equivalent circuit is presented in Figure 13.

Assuming now the values given in [30] for the negative resistance: R=-120 Ω, shunt capacitance C=0.7pF, while the series resistance $R_S$=3.5 Ω and the series inductance L=3.5 nH, let us analyse the frequency dependency of its input impedance from its $Q$ in between 5 GHz and 11 GHz. The quality factor of a tunnelling diode can be negative [30], while keeping the same classical definition. The evolution of the quality factor from values of -2 towards infinity can be seen in Fig. 16 (a)-(c) on a 3D Smith chart rendered with the constant normalized conductance (*g*) and susceptance (*b*) circles. The values of the normalized input admittance can be checked in each moment along the *g* and *b* circles also. While maintaining a negative input resistance, the device behaves capacitive with up to 8.3 GHz when *Q=0*, then from 8.3 GHz the device behaves inductive. Its *Q* becomes infinite in absolute value at 11 GHz, when its input resistance starts changing sign.

All these evolutions can be easily checked and computed with the 3D Smith chart without the need of further calculations.

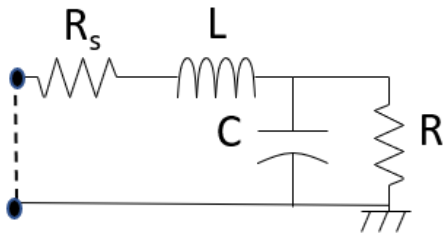

**FIGURE 15.** Model of the small signal equivalent circuit of a resonant tunnel diode consisting in a negative resistance *R*, shunt capacitance *C* and series resistance $R_s$ and series inductance *L*.

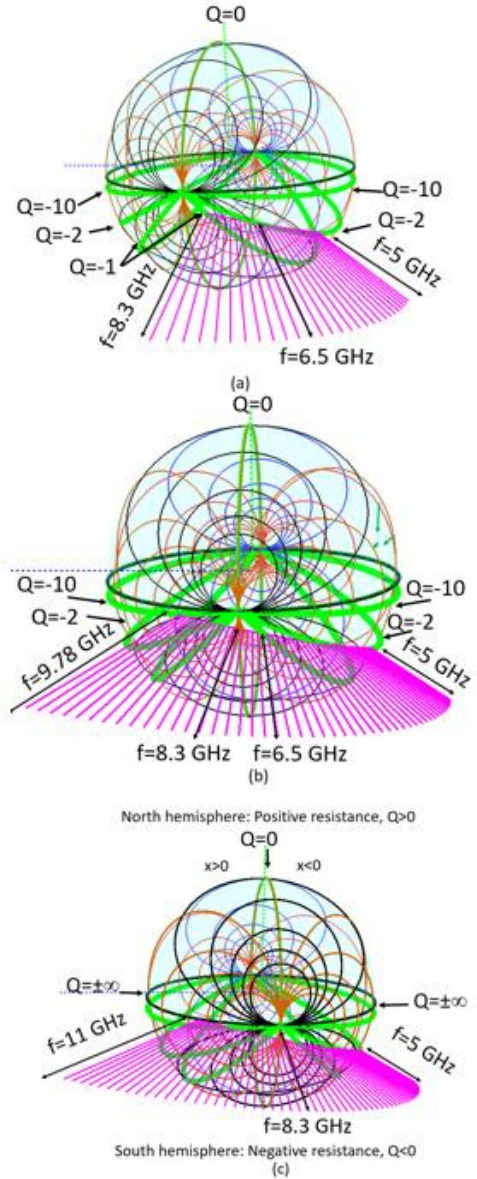

**FIGURE 16.** Reflection coefficient of the tunnel diode at a 50 Ohm port. (a) 5 GHz<*f*<8.3 GHz, *Q=-2* at 6.5 GHz, *Q=-1* at 8.3 GHz (b) 5 GHz<*f*<9.78 GHz, *Q= -10* at 9.78 GHz. (c) 5 GHz<*f*<11 GHz, |Q|=∞ at f=11 GHz.

### B. Single frequency point analysis

Let us now analyse the tunnel diode small signal equivalent circuit given in Fig. 17 at 2 GHz. ([2, pp 154]. Let us apply the 3D Smith chart implementation to compute the *Q* and input impedance in the points 1-2-3-4 from Fig. 17.

Employing a 2D Smith chart, this would be not possible to further since the negative resistance is thrown towards infinity. On the 3D Smith chart, we can start from *r=-1.25*, towards South pole, with *Q=0* (purely resistive). Thus, in point 1 we can read: *Q=0*, *r=-1.25* (or *g=-0.8*). Points1-2: We then move on *g=-0.8* circle to *b=1.9*. We can see that we touch the *Q=-1.94* circle, thus in point 2 we have *g=-0.8* and *b=1.9* (or *x=-0.45*). Points 2-3: we move on the *x=-0.45* constant circle adding r=0.02 and get to the point 3 where we intersect the $Q=-2.64$ circle. Thus, we obtain here $r=-0.17$ and $x=-0.45$. Points 3-4: We move on the $r=-0.17$ circle adding the $x=0.12$ value. In point 4 we can see directly that we cross the $Q=-1.94$ circle thus we obtain $z_{in}=-0.17+j0.33$, or in un-normalized coordinates: the input impedance becomes: $50*z_{in}=-8.5-j16.5$.

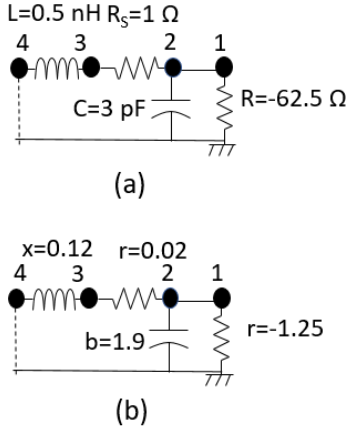

FIGURE 17. Small signal equivalent circuit of the tunnel diode ([2, p. 154] (a) real values at 2 GHz. (b) Normalized to 50 Ω impedance

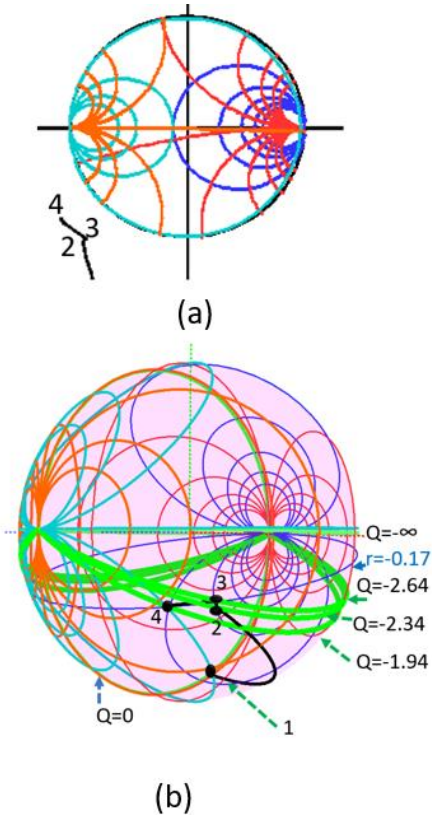

FIGURE 18. Input normalized impedance in the nodes 1-2-3-4 from Fig. 17 (a) on the Smith chart-impossible to plot (b) on the 3D Smith chart.

We can verify with ease the correctness of the approach by computing the input impedance mathematically. However, the 3D Smith chart implementations allowed us to follow step by step its development in different nodes, with no need of arithmetical manipulations, and, simply by changing the rendering, we were able to read the exact values stepwise. The appendix shows how the implementation can be used for a bandstop filter too in order to extract the equivalent circuit directly from the $S_{11}$ parameters intersections with constant $Q$ circles.

### IV. Conclusions

In this article we have proved, for the first time, that the constant $Q$ contours (nodal quality factors) (1) form circle arcs on a family of coaxal circles on the Smith chart. We provided, for the first time, (to the best of our knowledge) by means of bipolar equations, their explicit equations in terms of radius, circles centre-$Q$ value relationship, by solving their implicit equations. Further, we have proved that, while evaluated on the 3D Smith chart, the constant $Q$ contours represent semi-circles in the north hemisphere for positive resistance circuits, respectively semi-circles on the south hemisphere for negative resistance circuits, all cantered in 3D Smith chart centre. This simple, compact, and practical circle shaped property has enabled us to use these $Q$ semi-circles directly, in the reflection coefficients plane, for both passive and active circuits, for the direct $Q$ evaluations from measured S parameters. In the case of Vanadium Dioxide reconfigurable inductors temperature sensitivity analysis: the proposed methodology allowed us the multi-parameter extraction (inductance, $Q$, reflection coefficient) (Fig. 14) directly from the measured devices, simplifying the extraction procedures-and allowing us a fast evaluation of their performances directly from the measuring setup. In the case of negative resistance circuits, the proposed $Q$ visualization extended the use of constant Q contours for circuits with negative resistance too, impossible on a 2D Smith chart, exemplified here on tunnel diodes small signal equivalent circuits (Figs. 15-18).

### Appendix

Let us consider the parallel R, L, C (which can be the equivalent circuit of a bandstop filter) circuit present in Fig. 19. Supposing one would need to determine the values of the elements R, L, C in Fig. 19 that would fit a measured $S_{11}$ of a bandstop filter, whose equivalent circuit is completely determined by Fig. 19.: one can use the new frequency dependent $Q$ implementation.

By representing the $S_{11}$ with the second port grounded we get Fig. 20.

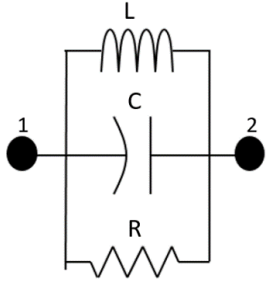

**FIGURE 19.** Bandstop filter equivalent circuit.

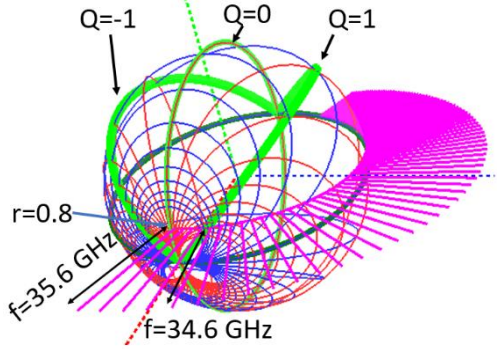

**FIGURE 20.** $S_{11}$ parameter of the resonant circuit in Fig. 17 with the second port grounded with unknown R,L,C.

The input admittance of the bandstop filter (Fig.19) can be computed with (14). At resonance the imaginary part is zero and (15) is fulfilled where $\omega_0$ is the angular resonance frequency, while $f_0$ the resonance frequency. The input admittance at resonance $Y_{110}$ becomes (16) and $Q$ defined in (13) becomes (17).

$$Y_{11} = \frac{1}{R} + j\left(\omega C - \frac{1}{\omega L}\right) \quad (14)$$

$$\omega_0 = \frac{1}{\sqrt{LC}}, f_0 = \frac{1}{2\pi\sqrt{LC}} \quad (15)$$

$$Y_{110} = \frac{1}{R} \quad (16)$$

$$\frac{Im\left(\frac{1}{Y_{11R}}\right)}{Re\left(\frac{1}{Y_{11R}}\right)} = 0 \quad (17)$$

In Fig. 20 we can easily determine $f_0$ and $R$: we check when $S_{11}$ crosses the $Q=0$ circle and read the values for the frequency and for the normalized resistance. We get $f_0 =35.6$ GHz and $r=0.8=R/50$, thus $R=400\Omega$.

Now let us compute (11) in a general form for the circuit given in Fig. 19:

$$Q = \frac{Im\left(\frac{1}{Y_{11}}\right)}{Re\left(\frac{1}{Y_{11}}\right)} = \frac{R}{\omega L} - RC\omega \quad (18)$$

Imposing now Q=1 we get:

$$\frac{R}{\omega_1 L} - RC\omega_1 = 1 \quad (19)$$

where $\omega_1=2*\pi f_1$ is the angular frequency for which Q=1 (and $f_1$ the frequency for which $Q=1$)

Getting back in Fig. 20 and checking where $S_{11}$ crosses the Q=1 circle we get $f_1= 34.6$ GHz. Now getting back to (15) and (19) we have:

$$35.6 GHz = \frac{1}{2\pi\sqrt{LC}} \quad (20)$$

$$\frac{400}{2\pi*34.6GHz*L} - 400C*2\pi*34.6GHz = 1 \quad (21)$$

Solving using Mathematica [31] numerically (20) and (21) we get one of the solutions: C=0.2 pF and L=0.1 nH. This enabled us to extract the equivalent circuit without any fitting procedure.

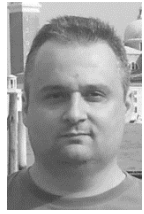

**Victor Asavei** received the Ph.D. degree in Computer Science and Information Technology from University Politehnica of Bucharest in 2011. He is an Assistant professor (Lecturer) at University Politehnica of Bucharest, Faculty of Automatic Control and Computers, Computer Science and Engineering Department. His current teaching and research domains are in the areas of Real Time Computer Graphics and GPGPU (General Purpose computing on Graphics Processing Units).

Dr. Asavei has co-authored approximately 60 papers and 4 books in the fields of Computer Graphics, Distributed Computing, Software Engineering and Medical ITs and has participated in numerous national and international research projects.

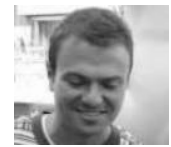

**Andrei A. Muller** was born in Bucharest, Romania. He completed his studies in Mobile and Satellite Communications and further a PhD from the Politehnica Bucharest in microwaves engineering. Andrei is a recipient of the Gheorghe Cartianu prize of the Romanian Academy of Science (2013) for the article "A 3D Smith chart Riemann Sphere for Active and Passive Microwave Circuits-IEEE MWCL, 2011". He is a Senior Member of IEEE since 2018, and an Outstanding Associate Editor of the IEEE Access in 2017 and 2018. Being an ex Marie-Curie Fellow in i-team UPV Valencia, Spain, Andrei is nowadays a Scientist at Nanolab, EPFL Lausanne since September 2017. His current interests include CAD and smart materials for RF and microwaves engineering.

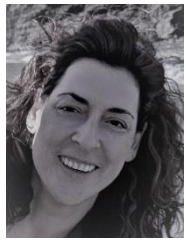

**Esther Sanabria-Codesal.** Esther Sanabria-Codesal received the Ph.D. degree in geometry and topology from the University of Valencia, Valencia, Spain, in 2002. She is an Associate Professor of the Department of Applied Mathematics at Universitat Politècnica de València, Valencia. Her current research interests include theory of singularities applied to geometry and mathematical modelling based on graphs. Dr. Sanabria-Codesal has participated in numerous research projects and conferences.

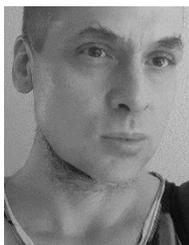

**Alin Moldoveanu** is Professor (teaching Software Engineering and Virtual Reality) and Vice-dean (in charge of the master studies) at this same faculty that he completed as valedictorian, 20 years ago – Faculty of Automatic Control and Computers, from POLITEHNICA University of Bucharest.

He's teaching software engineering at bachelor level and virtual and augmented reality at masters, and focusing on applied research in virtual and augmented reality (exploring and applying immersion, sensory substitution, and distorted reality), eHealth (assistive and rehabilitative solutions, prevention of hospital acquired infections) and eLearning & eCulture (mixed-reality campuses and cultural environments). He is the director or responsible for many national or European research projects in these areas, such as Sound of Vision, TRAVEE, HAI-OPS, Lib2Life. List of projects and publications available at: https://cs.pub.ro/index.php/people/userprofile/alin_moldoveanu

Dr. Moldoveanu's research works received several prestigious prizes, such as Best *"Tech for Society" Horizon 2020 project*, awarded by EC through Innovation Radar, at ICT 2018 – received by Sound of Vision, where he acted as technical coordinator and UPB team responsible.

**Adrian M. Ionescu**